\documentclass[11pt,twoside]{article}
\usepackage{./asp2010}

\resetcounters

\begin{document}

\title{A New Approach to Detailed Structural Decomposition: \\
Kicked-up
Disk Stars in Andromeda's Halo?} 
\author{Claire E. Dorman,$^1$ Lawrence M. Widrow,$^2$ Puragra
  Guhathakurta,$^1$ \\
 and the PHAT collaboration
\affil{$^1$Department of Astronomy \& Astrophysics, University of
  California at Santa Cruz, 1156 High St., Santa Cruz, CA 95064}
\affil{$^2$Department of Physics, Engineering Physics \& Astronomy,
  Queen's University, Kingston, Ontario, Canada}}

\begin{abstract}
We characterize the bulge, disk, and halo subcomponents in the
Andromeda galaxy (M31) over the radial range $0.4~{\rm kpc}<R_{\rm
  proj}<225$ kpc. The cospatial nature of these subcomponents renders
them difficult to disentangle using surface brightness (SB)
information alone, especially interior to $\sim 20$ kpc. 
Our new decomposition technique combines information
from the luminosity function (LF) of over $1.5$ million bright $(20 <
m_{\rm F814W} < 22)$ stars from  the Panchromatic Hubble Andromeda
Treasury (PHAT) survey, radial velocities of
over $5000$  red giant branch stars in the same magnitude range from
the Spectroscopic and Photometric Landscape of Andromeda's Stellar Halo
(SPLASH) survey, and integrated $I$-band SB profiles from
various sources. We use an affine-invariant Markov chain Monte Carlo algorithm to fit an
appropriate toy model to these three data sets. The bulge, disk, and halo
SB profiles are modeled as a S\'ersic, exponential, and cored power-law,
respectively, and the LFs are modeled as broken
power-laws.  We find that the number of stars with a disk-like LF is
$\sim 5\%$ larger than the number in the dynamically cold component,
suggesting that some stars born in the disk have been dynamically
heated to the point that they are kinematically indistinguishable from
halo members. This is the first kinematical evidence for a ``kicked-up
disk'' population in the stellar halo of M31. The fraction of kicked-up disk stars is
consistent with that found in simulations. See \citet{dor13} for more
information. 

\end{abstract}

\section{Introduction}
Structural decomposition of a spiral galaxy is typically done by fitting a
sum of a  model bulge, disk, and halo to the galaxy's
surface brightness (SB) profile. However, this method
suffers from degeneracies in the best-fit profiles, particularly in
the inner ($5 < R_{\rm proj} < 20$ kpc) regions where the three
components are cospatial. In this work, we introduce a new
decomposition technique that uses resolved stellar population 
data as additional constraints in a SB decomposition of Andromeda, the
nearest large spiral galaxy.

\section{Simultaneous fit to three data sets}
We fit a simple toy model simultaneously to three data sets. First, we
use I-band surface brightness measurements in 637 fields between 0.4
and 225 kpc in projected galactocentric radius from
\cite{cou11,gil12}. We model the SB map as the sum
of a S\'ersic bulge, exponential disk, and power-law halo, each with
their own ellipticity parameters but the same major axis position
angle. 

Second, we use the $m_{\rm F814W}$ luminosity function (LF) from the
Panchromatic Hubble
Andromeda Treasury (PHAT; \citet{dal12}) survey in each of
14 small spatial subregions in the inner 20 kpc of the galaxy. We only
use the magnitude range surrounding the tip of the red giant branch
(TRGB): $20 < m_{\rm F814W} < 22$, which is bright enough that it is
essentially 100\% complete throughout the survey area. We exclude from
the LF stars
in regions with $A_v > 1.0$. We model the LF as
the sum of three individual LFs, one each for the bulge, disk, and
halo. Each individual LF is parameterized as a broken power-law,
so that the slope has the freedom to change at the TRGB associated
with the population in a given component. The shapes of the model bulge and
halo LFs are required to be constant (though their normalizations can
change with position), but the shape parameters of the model disk LF
are allowed to change with radius in the plane of the disk. 

Finally, we fit our model to the fraction of stars kinematically associated with the
disk (the ``disk fraction'') in each subregion. Figure~1 illustrates
the measurement of the disk fraction in 4
representative subregions. In each subregion, we use a Markov chain
Monte Carlo (MCMC) algorithm to fit a sum of two Gaussian
distributions, representing the dynamically cold disk (gray dashed
curve) and warmer spheroid (dotted curve) to the line-of-sight velocity distribution of red
giant branch stars from the Keck/DEIMOS SPLASH survey. The measured
disk fraction in a subregion is given by the ratio of the area under the
disk Gaussian to the area under the sum of the disk and spheroid
Gaussians. See \citet{dor12} for a more detailed description. 

The model disk fraction in a given subregion is determined by the model SB
decomposition in that subregion. However, while the disk
fraction measurements constrain the fraction of stars sampled by the
SPLASH survey that belong to the disk, the SB model predicts the
fraction of integrated light contributed by the disk. To compare the
two, we convert the SB model prediction to a disk fraction in
units of SPLASH-sampled star counts using the empirical SPLASH
selection function in each subregion and the model disk luminosity
function. 

We define a likelihood function --- approximately the sum of the
goodness-of-fit $\chi^2$ parameters between the model and data for the
SB, LF, and disk fractions --- and sample it using the MCMC
sampler {\tt emcee} \citep{for12} to measure the probability
distributions of each of the 32 parameters in the toy model. We can
extract the median value and $1\sigma$ uncertainty on each parameter
as well as measure covariances between model parameters. 

\begin{figure}[!ht]
\plotone{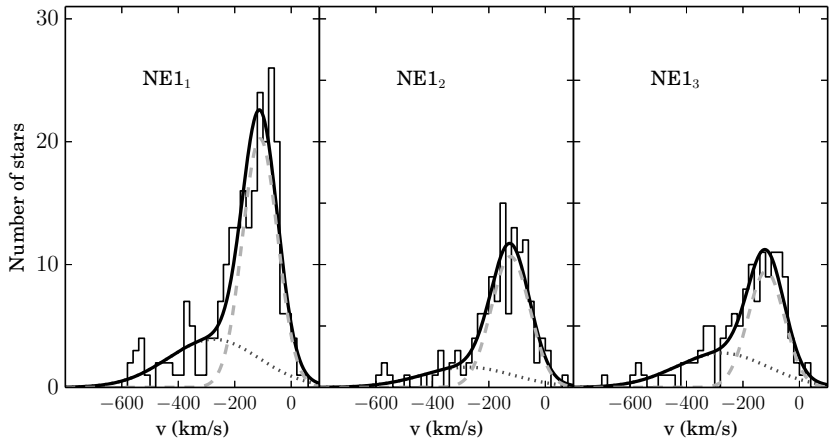}
\caption{Kinematical decomposition in three representative subregions
  in a reginon centered on the major axis of M31 around $8$ kpc from
  the center.  In each subregion, the disk fraction is
the ratio of the areas of the spheroid and total distributions.}
\end{figure}

\section{Kicked-up disk stars in the halo?}

The fits to the SB, kinematics and luminosity functions in
representative subregions are shown in Figures 2 and 3. The LF and SB
are well fit, but the right-hand panel of Figure 2 shows that the
kinematics are not. Here, the kinematically-derived cold fraction is
plotted against the best-fit model disk fraction. The model
systematically overpredicts the cold contribution by $\sim 5\%$ on
average. The inability to simultaneously fit the three data sets is a
sign of tension between the simple model and the data. The tension
would be reduced if some fraction of the population with a disk-like LF (that is,
stars that originated in the disk) had been dynamically heated (that is, had
halo-like kinematics). In this scenario, the fraction of dynamically
hot disk stars (the ``kicked-up disk'') is consistent with that predicted by cosmological
simulations \citep{pur10, tis13}. 

Of course, our decomposition depends on our choice of model. For
the sake of completeness we perform decompositions with modified SB
models, including a broken exponential disk; two exponential disks;
two S\'ersic bulges; a single bulge with distinct inner and outer
S\'ersic indices; and a S\'ersic halo. None can simultaneously fit all
three data sets without  invoking a kicked-up disk. 

A more detailed discussion of this work can be found in
\citet{dor13}. 

\begin{figure}[!ht]
\plottwo{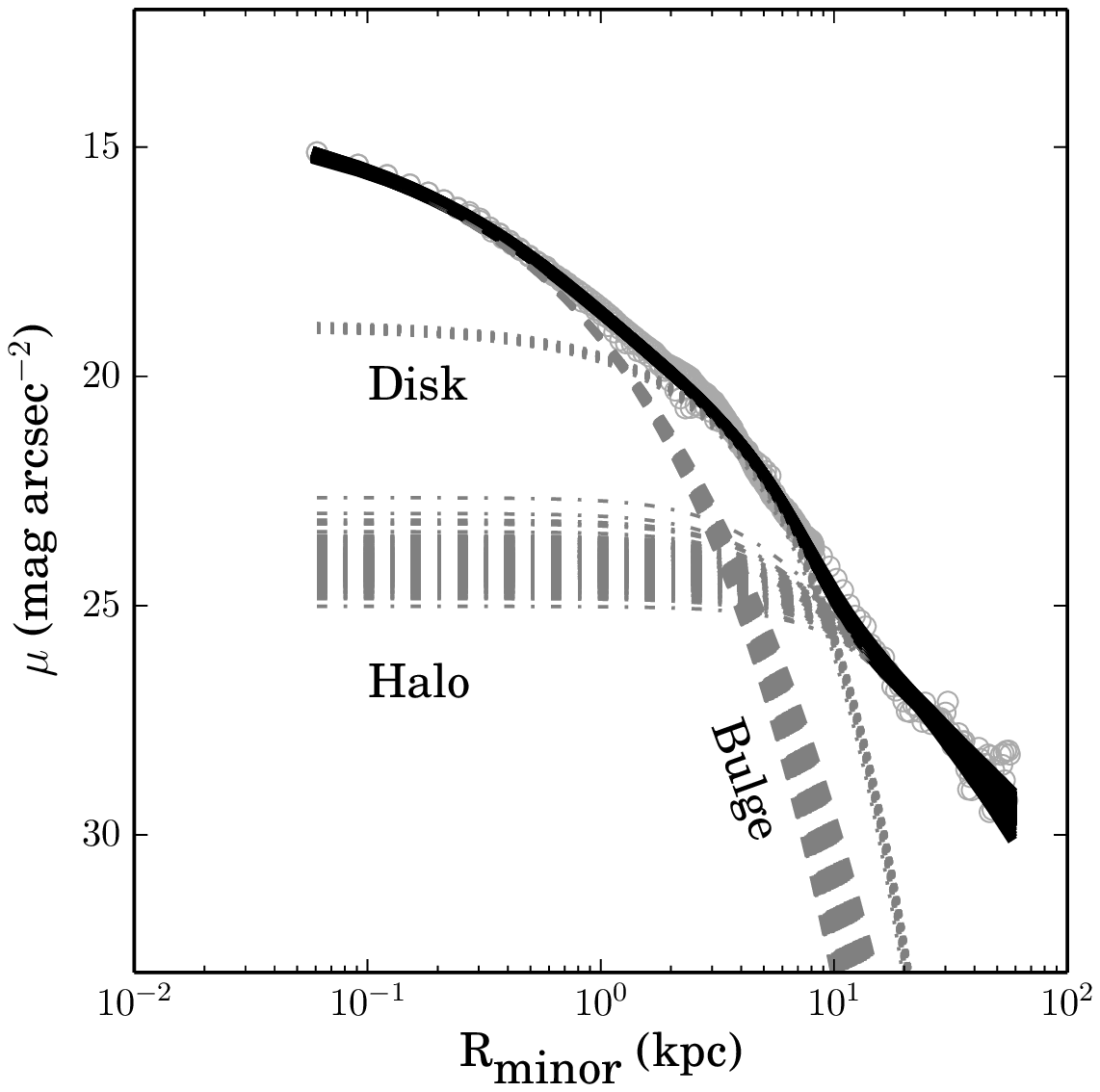}{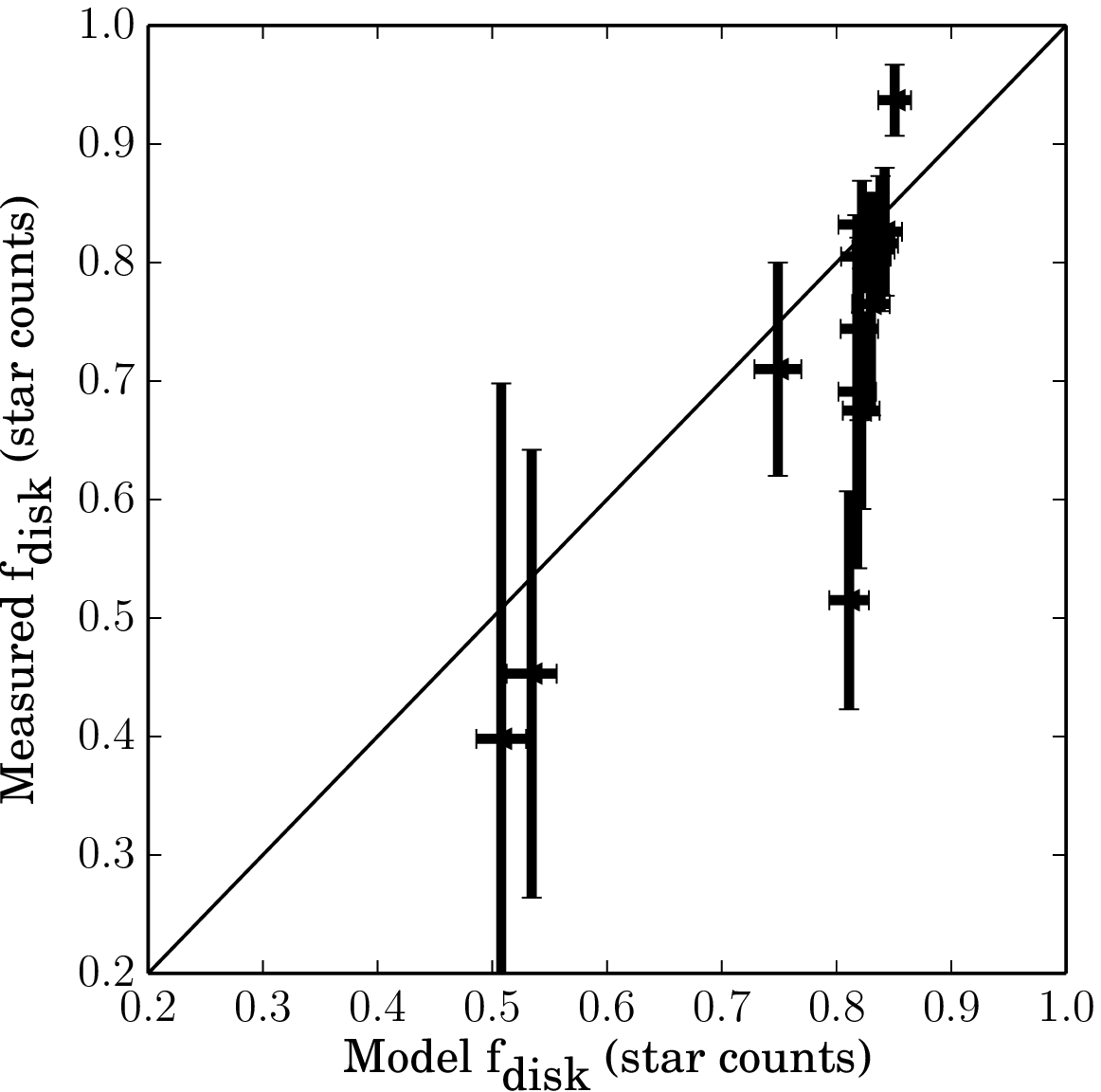}
\caption{{\em Left:} Minor axis projection of the fit to the SB
  profile. For each profile (disk, bulge, and halo), the distribution
  of lines represents the entire range of allowed profiles, rather than the $1\sigma$
uncertainty. The set of black lines shows the total model, and the
gray dots (barely visible) the observed minor-axis SB profile. {\em
  Right:} Measured vs. best-fit model disk fraction, as computed in
units of SPLASH star counts. The model disk fractions are
systematically higher than the measured disk fractions, suggesting
that some fraction of the stars with a disklike LF may be dynamically
hot.}
\end{figure}

\begin{figure}[!ht]
\plotone{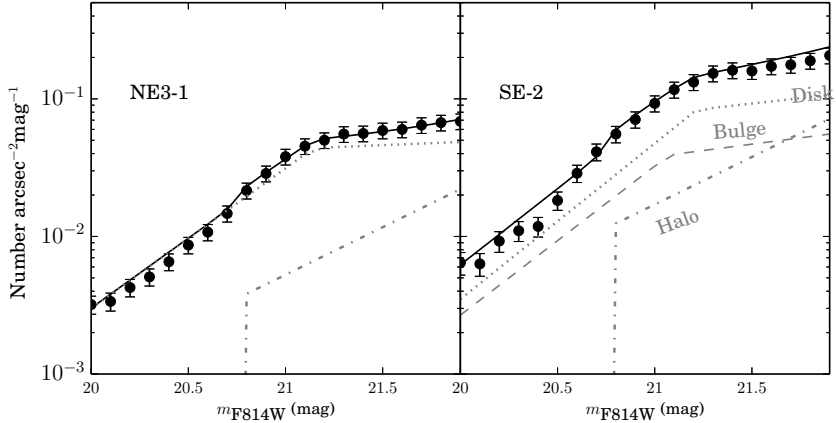}
\caption{Fit to the observed LF from the PHAT survey in two
  representative subregions: one on the major axis {\em (left)} and one on
  the minor axis {\em (right)}. Circles with error bars show the observed
  LF; dotted, dashed, and dot-dashed lines show the model disk, bulge
  and halo LFs, respectively. The solid line represents the sum of the
  three model LFs. The disk contributes more stars than the bulge and
  halo in every subregion.}
\end{figure}

\acknowledgements The authors acknowledge NSF grants AST-0607852 and
AST-1010039, NASA grant HST-GO-12055, and a NSF Graduate Research
Fellowship.

\bibliography{cdorman}

\begin{thebibliography}{}
\expandafter\ifx\csname natexlab\endcsname\relax\def\natexlab#1{#1}\fi
\expandafter\ifx\csname url\endcsname\relax
  \def\url#1{\texttt{#1}}\fi
\expandafter\ifx\csname urlprefix\endcsname\relax\def\urlprefix{URL }\fi
\providecommand{\eprint}[2][]{\url{#2}}

\bibitem[{{Courteau} et~al.(2011){Courteau}, {Widrow}, {McDonald},
  {Guhathakurta}, {Gilbert}, {Zhu}, {Beaton}, \& {Majewski}}]{cou11}
{Courteau}, S., {Widrow}, L.~M., {McDonald}, M., {Guhathakurta}, P., {Gilbert},
  K.~M., {Zhu}, Y., {Beaton}, R.~L., \& {Majewski}, S.~R. 2011, \apj, 739, 20.
  \eprint{1106.3564}

\bibitem[{{Dalcanton} et~al.(2012){Dalcanton}, {Williams}, {Lang}, {Lauer},
  {Kalirai}, {Seth}, {Dolphin}, {Rosenfield}, {Weisz}, {Bell}, {Bianchi},
  {Boyer}, {Caldwell}, {Dong}, {Dorman}, {Gilbert}, {Girardi}, {Gogarten},
  {Gordon}, {Guhathakurta}, {Hodge}, {Holtzman}, {Johnson}, {Larsen}, {Lewis},
  {Melbourne}, {Olsen}, {Rix}, {Rosema}, {Saha}, {Sarajedini}, {Skillman}, \&
  {Stanek}}]{dal12}
{Dalcanton}, J.~J., {Williams}, B.~F., {Lang}, D., {Lauer}, T.~R., {Kalirai},
  J.~S., {Seth}, A.~C., {Dolphin}, A., {Rosenfield}, P., {Weisz}, D.~R.,
  {Bell}, E.~F., {Bianchi}, L.~C., {Boyer}, M.~L., {Caldwell}, N., {Dong}, H.,
  {Dorman}, C.~E., {Gilbert}, K.~M., {Girardi}, L., {Gogarten}, S.~M.,
  {Gordon}, K.~D., {Guhathakurta}, P., {Hodge}, P.~W., {Holtzman}, J.~A.,
  {Johnson}, L.~C., {Larsen}, S.~S., {Lewis}, A., {Melbourne}, J.~L., {Olsen},
  K.~A.~G., {Rix}, H.-W., {Rosema}, K., {Saha}, A., {Sarajedini}, A.,
  {Skillman}, E.~D., \& {Stanek}, K.~Z. 2012, \apjs, 200, 18

\bibitem[{{Dorman} et~al.(2012){Dorman}, {Guhathakurta}, {Fardal}, {Lang},
  {Geha}, {Howley}, {Kalirai}, {Bullock}, {Cuillandre}, {Dalcanton}, {Gilbert},
  {Seth}, {Tollerud}, {Williams}, \& {Yniguez}}]{dor12}
{Dorman}, C.~E., {Guhathakurta}, P., {Fardal}, M.~A., {Lang}, D., {Geha},
  M.~C., {Howley}, K.~M., {Kalirai}, J.~S., {Bullock}, J.~S., {Cuillandre},
  J.-C., {Dalcanton}, J.~J., {Gilbert}, K.~M., {Seth}, A.~C., {Tollerud},
  E.~J., {Williams}, B.~F., \& {Yniguez}, B. 2012, \apj, 752, 147.
  \eprint{1204.4455}

\bibitem[{{Dorman} et~al.(2013){Dorman}, {Widrow}, {Guhathakurta}, {Seth},
  {Foreman-Mackey}, {Bell}, {Dalcanton}, {Gilbert}, {Skillman}, \&
  {Williams}}]{dor13}
{Dorman}, C.~E., {Widrow}, L.~M., {Guhathakurta}, P., {Seth}, A.~C.,
  {Foreman-Mackey}, D., {Bell}, E.~F., {Dalcanton}, J.~J., {Gilbert}, K.~M.,
  {Skillman}, E.~D., \& {Williams}, B.~F. 2013, \apj ~(in press)

\bibitem[{{Foreman-Mackey} et~al.(2012){Foreman-Mackey}, {Hogg}, {Lang}, \&
  {Goodman}}]{for12}
{Foreman-Mackey}, D., {Hogg}, D.~W., {Lang}, D., \& {Goodman}, J. 2012.
  \eprint{1292.3665}

\bibitem[{{Gilbert} et~al.(2012){Gilbert}, {Guhathakurta}, {Beaton}, {Bullock},
  {Geha}, {Kalirai}, {Kirby}, {Majewski}, {Ostheimer}, {Patterson}, {Tollerud},
  {Tanaka}, \& {Chiba}}]{gil12}
{Gilbert}, K.~M., {Guhathakurta}, P., {Beaton}, R.~L., {Bullock}, J., {Geha},
  M.~C., {Kalirai}, J.~S., {Kirby}, E.~N., {Majewski}, S.~R., {Ostheimer},
  J.~C., {Patterson}, R.~J., {Tollerud}, E.~J., {Tanaka}, M., \& {Chiba}, M.
  2012, \apj, 760, 76. \eprint{1210.3362}

\bibitem[{{Purcell} et~al.(2010){Purcell}, {Bullock}, \& {Kazantzidis}}]{pur10}
{Purcell}, C.~W., {Bullock}, J.~S., \& {Kazantzidis}, S. 2010, \mnras, 404,
  1711. \eprint{0910.5481}

\bibitem[{{Tissera} et~al.(2013){Tissera}, {Beers}, {Carollo}, \&
  {Scannapieco}}]{tis13}
{Tissera}, P., {Beers}, T., {Carollo}, D., \& {Scannapieco}, C. 2013, ArXiv
  e-prints. \eprint{1309.3609}

\end{thebibliography}
\end{document}